\documentclass[12pt,preprint]{aastex}

\shorttitle{Accretion in a Brown Dwarf of  R~CrA}
\shortauthors{Barrado y Navascu\'es  et al.}

\begin{document}

\def\hal{${H\alpha}$ }
\def\na{\ion{Na}{1}\ }
\def\pot{\ion{K}{1}\ }
\def\li{\ion{Li}{1}\ }
\def\ca{\ion{Ca}{2}\ }
\def\kms{km s$^{-1}$\ }
\def\msun{M$_\odot$\ }
\def\lsun{L$_\odot$\ }
\def\mj{M$_J$\ }
\def\teff{T$_{e\! f\! f}$~}
\def\gv{{\it g}~}
\def\vsini{{\it v}~sin{\it i}~}
\def\vrad{v$_{\it rad}$~}
\def\lbol{L$_{bol}$~}
\def\lhal{L$_{\H\alpha}$ }
\def\eqwhal{EW$_{H\alpha}$ }
\def\logten{$log_{10}$ }
\def\jh{$J$-$H$ }
\def\jk{$J$-$K$ }
\def\jks{$J$-$K_S$ }
\def\aj{A$_J$ }
\def\aks{A$_{Ks}$ }
\def\bcj{BC$_J$ }

\title{\normalsize{Accretion and Outflow in the Substellar Domain: \\
 Magellan Spectroscopy of LS-RCrA~1}}

\author{David Barrado y Navascu\'es}
\affil{Laboratorio de Astrof\'{\i}sica Espacial y F\'{\i}sica Fundamental,
INTA, P.O. Box 50727, E-2808 Madrid, SPAIN}
\email{barrado@laeff.esa.es}
\author{Subhanjoy Mohanty}
\affil{Harvard-Smithsonian Center for Astrophysics,
 60 Garden St., Cambridge, MA 02138, U.S.A.}
\email{smohanty@cfa.harvard.edu}
\and
\author{Ray Jayawardhana}
\affil{Department of Astronomy, University of Michigan, 
830 Dennison Building, Ann Arbor, MI 48109, U.S.A.}
\email{rayjay@umich.edu}

\begin{abstract}
\small{We present low-, medium-, and high-resolution optical spectra, obtained with the Magellan Baade 6.5m telescope, of LS-RCrA~1, a late-type object identified recently by Fern\'andez and Comer\'on (2001) in the R Coronae Australis star-forming region.  We confirm both pre-main sequence status and membership in the R~CrA region for this object, through the detection of \li, presence of narrow \pot indicative of low gravity, and measurement of radial velocity.  The H$\alpha$ emission profile is very broad, with a 10\% full width of 316 \kms at high-resolution, implying the presence of ongoing accretion.  Our spectra also exhibit many forbidden emission lines indicative of mass outflow, in agreement with the Fern\'andez and Comer\'on results.  We derive a spectral type, independent of extinction, of M6.5$\pm$0.5 IV. Using new  2MASS near-infrared photometry, no  significant NIR excess is found.  Our optical veiling measurements yield a mass accretion rate of 10$^{-10}$ $\lesssim$ \.{M} $\lesssim$ 10$^{-9}$ {\msun}yr$^{-1}$.  The presence of prominent outflow signatures at these low accretion rates is initially puzzling.  We consider, and discard as improbable, the possibility that these signatures arise in a line-of-sight Herbig-Haro knot unassociated with LS-RCrA~1 itself.  However, if LS-RCrA~1 possesses a nearly edge-on disk, a natural outcome would be the enhancement of any outflow signatures relative to the photosphere; we favor this view.  A low accretion/outflow rate, combined with an edge-on orientation, is further supported by the absence of high-velocity components and any significant asymmetries in the forbidden lines.  An edge-on geometry is also consistent with the lack of NIR excess in spite of ongoing accretion, and explains the relatively large \hal 10\% width compared to other low-mass objects with similar accretion rates.  Through comparison with the latest synthetic spectra, we infer \teff $\approx$ 2700$\pm$100K, somewhat lower than the previous estimate (2900$\pm$200K).  Theoretical evolutionary tracks then imply an age of $\sim$20 Myr (as derived from \teff and luminosity) or $\sim$8 Myr (\teff vs. gravity) for LS-RCrA~1. This last value is consistent with the estimated age of other T Tauri stars in R~CrA ($\lesssim$ 10 Myr), and it is substantially less than the $\sim$ 50 Myr derived previously. Therefore,  LS-RCrA~1 indeed appears sub-luminous  relative to expectations for an R~CrA member.  By comparing its position on the H-R diagram with that of other similarly accreting low-mass objects, we show that accretion-induced effects are unlikely to account for its faintness.  We suggest instead that LS-RCrA~1 posesses a nearly edge-on disk and its photosphere is seen predominantly in scattered light, making it appear much fainter (and older) than it really is.  The ease with which such a disk simultaneously explains all the puzzling aspects of LS-RCrA~1 -- sub-luminosity, unusually prominent outflow signatures without high-velocity components or asymmetries, very broad \hal, lack of NIR excess combined with accretion -- makes its presence a strong possibility.  Finally, the surface gravity  and \teff estimates, combined with the latest evolutionary tracks, indicate a mass of $\sim$0.07 or 0.035$\pm$0.010 \msun (depending on the \teff scale), i.e., at or below the substellar boundary.  Our results, together with those of Fern\'andez and Comer\'on, imply that young brown dwarfs can not only harbor accretion disks but also generate jets/outflows analogous to those in higher mass classical T Tauri stars.  This is further evidence of a common formation mechanism for stars and brown dwarfs.}
\end{abstract}


\section{Introduction}
The origin and early evolution of very low mass objects is a topic of significant current interest and research.  Theorists debate whether brown dwarfs form via fragmentation and collapse of molecular clouds (e.g., Padoan \& Nordlund 2003) or as stellar embryos ejected from newborn multiple systems (Reipurth \& Clarke 2001; Bate, Bonnell \& Bromm 2002). Recent observations have provided compelling evidence that young sub-stellar objects undergo a T Tauri-like accretion phase similar to their stellar counterparts (Jayawardhana, Mohanty \& Basri 2002, 2003; Barrado y Navascu\'es et al. 2002, 2003a; Barrado y Navascu\'es \& Mart\'{\i}n 2003) and harbor infrared excesses consistent with dusty disks (Muench et al. 2001; Natta et al. 2002; Jayawardhana et al. 2003; Liu, Najita \& Tokunaga 2003). The analogy between (solar-mass) T Tauri stars and young brown dwarfs would be strengthened further if some of the latter also exhibit evidence of mass outflows.  

In that context, a very low mass object recently identified by Fern\'andez and Comer\'on (2001, hereafter F\&C01) in the R Coronae Australis region, and named LS-RCrA~1, is particularly intriguing.  Their low-resolution (R$\sim$500) optical spectrum revealed permitted emission lines with equivalent widths typical of accreting T Tauri stars, as well as forbidden lines usually associated with mass-loss. The $K$-band spectrum shows the H$_2$ line at 2.12$\mu$m in emission and CO bands, but no other discernible features. H$_2$ is a tracer of molecular outflows and is frequently seen in accreting Class~I sources (Greene \& Lada 1996).  On the basis of these observations, F\&C01 suggested that LS-RCrA~1 is undergoing very intense accretion and mass-loss.  However, they were puzzled by its photometric properties, which suggested (when compared to theoretical evolutionary models) that the object was either much older or much fainter than expected for an R~CrA member.  
 
Here we considerably enhance the available data for LS-RCrA~1, by presenting low-, medium-, and high-resolution optical spectra obtained at Magellan and accurate near-infrared (NIR) photometry from 2MASS.  Using these data, we derive membership information, spectral type, optical veiling and extinction independently, and effective temperature, mass and age with the aid of theoretical models.  We compare our results to those of F\&C01.  

\section{Observations and Data Reduction}
We obtained spectra of LS-RCrA~1 at low-resolution (March 9, 2003) and at medium-resolution (March 10 \& 11, 2003) with the Boller \& Chivens (B\&C) spectrograph on the Magellan Baade 6.5-meter telescope.  The spectral resolutions were R=620 and R=2600, as measured in the comparison arcs, using the 300 l/mm and 1200 l/mm gratings.  The spectral coverages were  5200--9600 \AA{ } and 6200--7800 \AA~ respectively.  We collected several consecutive observations each night.  Each of them was  600 seconds for the first night and 900-1200 seconds for the last two nights, totalling 2400 sec, 5100 sec and 3500 sec.  The data were reduced using standard techniques within IRAF\footnote{IRAF is distributed by National Optical Astronomy Observatories, which is operated by the Association of Universities for Research in Astronomy, Inc., under contract to the National Science Foundation, USA}.  We also obtained high-resolution spectra (May 9, 2003) using the Magellan Inamori Kyocera Echelle (MIKE) spectrograph on Baade (Bernstein et al. 2002). The spectral resolution was R=19,000, with complete spectral coverage between 3640 and 8715 \AA.

\section{Results and Discussion}
\subsection{Spectral Type, Pre-Main-Sequence Status and Membership}
On the same nights when our LS-RCrA~1 low- and medium-resolution spectra were obtained, we also observed an array of stars of spectral types K7--M9 and luminosity classes III, IV and V.  The spectral type of LS-RCrA~1 was first estimated through direct comparison with these templates, over a spectral region including the gravity-sensitive potassium resonance doublet (\pot, $\sim$ 7700 \AA; Fig. 1). We see that the M6 dwarf gives an adequate fit to the continuum slope, but the \pot lines in it are vastly stronger than in LS-RCrA~1.  This is consistent with the latter being a young pre-main sequence (PMS) object, with gravity significantly lower than in a Main Sequence star of similar type; we revisit this issue below.  The M6III and M6--M7IV templates are good matches to the overall spectrum in this region.  However, the \pot lines in the M6III template are somewhat narrower than in LS-RCrA~1, suggesting a slightly higher gravity in the latter (again as expected for a low-mass PMS object).  Finally, the strength of the doublet in LS-RCrA~1 is comparable to that in the M6--M7IV templates, but the overall flux in the doublet and overlying continuum appears higher in our object than in these templates.  Given the presence of strong accretion-like \hal in LS-RCrA~1 (\S 3.4), we are led to suspect that this might be due to optical veiling caused by excess emission from an accretion shock.  In \S 3.2, we confirm such veiling. For now, we conclude that, while the medium-resolution spectra indicate a spectral type of M6--M7IV, veiling makes this uncertain; it is advisable to check the spectral type through some other means.  White \& Basri (2003; hereafter WB03) have shown that at the reddest optical wavelengths, the veiling is modest enough to determine spectral type to within $\pm$0.5 subclasses even in so-called continuum T Tauri stars (where veiling creates a largely featureless spectrum in the blue).  Following these authors, we rederive the spectral type by comparing the temperature sensitive TiO bands around 8440 \AA~ in LS-RCrA~1 to those in non-accreting Chameleon I PMS objects of various spectral types, at high-resolution (the Cha objects were observed by us in a previous Magellan run with the same setup).  As Fig. 2 shows, M6.5 a provides a very good match: though there is broad permitted OI emission in LS-RCrA~1 around 8445\AA~ (expected in an accretor; see Jayawardhana, Mohanty \& Basri (2003), hereafter JMB03), the bandheads at 8432, 8442 and 8451 \AA~ as well as the surrounding continuum are largely unaffected, enabling adequate fits for spectral type determination.  We thus adopt M6.5$\pm$0.5 for this object (consistent with the F\&C01 estimate of M6--M7).  

As mentioned above, the weakness of the \pot doublet in LS-RCrA~1, compared to a field dwarf of similar spectral type, attests to its low gravity and hence PMS status.  Moreover, we also detect Lithium at 6707 \AA~ in our medium-resolution spectrum\footnote{the high-resolution spectra of this faint, red object are too noisy in the relatively blue \li region to detect the line.} (see Fig. 3, bottom), though the line appears relatively weak, presumably due to continuum veiling as well as proximity to the [SII] emission.  This is further evidence of PMS status: field M dwarfs do not exhibit \li, having depleted it fairly early during their descent onto the Main Sequence.  

Finally, we derive a \vsini of 18 $\pm$ 3 \kms for LS-RCrA~1, and a radial velocity (\vrad) of +2 $\pm$ 3 \kms, both through cross-correlation of its high-resolution Magellan spectra with that of the M6 V standard Gl 406 (the faintness of LS-RCrA~1, combined with its moderately rapid rotation rate, precluded a more accurate \vrad estimate).  For 12 earlier type T Tauri stars in the CrA complex, Neuha\"user et al. (2000; hereafter N00) have found velocities ranging from 0 to -5 \kms (mean -2.6 \kms).  Our \vrad falls slightly outside this range, if our uncertainties are ignored, but is consistent with the N00 values when our errors are included.  Moreover, considering the spread in velocities found by N00 in their relatively small sample, a velocity of +2 \kms is not strikingly aberrant either.  In short, our \vrad appears compatible with the known values for the CrA complex. Taken together with the clear spectroscopic signatures mentioned above of PMS status, this  indicates that LS-RCrA~1 is a bona-fide PMS member of the R~CrA star-forming region (specifically, PMS status is indisputable; cluster membership, while suggested by our \vrad, needs to be established more stringently with a better \vrad determination).  

\subsection{Veiling and Extinction}
We estimate the optical continuum veiling, due to excess emission from the accretion shock, by comparing the strength of several TiO molecular bands in our low and medium resolution spectra of LS-RCrA~1 to those of templates of similar spectral type and luminosity type.  We establish that this veiling (defined as $r$($\lambda$) = $F(\lambda)_{excess}/F(\lambda)_{photosphere}$) is strongly wavelength-dependent, rapidly decreasing with longer wavelength: we obtain $r_{6200}$ = 1.00$\pm$0.20, $r_{6750}$ = 0.25$\pm$0.10, and $r_{7150}$ = 0.15$\pm$0.05.  We note that F\&C01 found substantial veiling even at $\sim$ 2.2$\mu$m (from the filling in of K-band absorption features): $r_K$ = 0.7-2.0.  However, this is not at odds with our results.  The K-band excess, as F\&C01 discuss, is expected to arise from either dust continuum emission or {\it line}-emission from the accretion shock, not continuum emission from the shock region, while we measure the latter in the optical.  We address the infrared (IR) veiling after calculating the extinction. 

We derive extinction using \jh color (like F\&C01): this should be much less affected by accretion-related continuum emission than optical colors, be relatively unaffected by any IR disk emission (which gains prominence at still longer wavelengths), and also be relatively insensitive to uncertainties in the extinction estimate.  Here we use the more recent and accurate 2MASS photometry for LS-RCrA~1 (All Sky Release, Cutri et al., 2003), which gives \jh=0.65.  Luhman (1999) argued that the intrinsic NIR colors of M-type PMS objects are dwarf-like.  We thus compare the observed \jh of LS-RCrA~1 to that of unreddened field dwarfs of similar spectral type; as a check, we also compare to negligibly reddened similar type PMS objects in the Pleiades and Alpha Per.  For M6.5 dwarfs, Leggett (1992) quotes an average \jh of 0.59 (on the CIT photometry system), while for Pleiades M6.5 objects, Zapatero Osorio, Mart\'{\i}n \& Rebolo (1997; hereafter ZMR97) find 0.60--0.85 (on the UKIRT system).  Alpha Per M6.5 members discovered by Stauffer et al. (1999), have a \jh in the range 0.66--0.75 in the 2MASS system (Barrado y Navascu\'es et al. 2001).  Transforming the dwarf and Pleiades colors to the 2MASS system (using Carpenter, 2001), and employing the reddening law of Rieke \& Lebofsky (1985), yields \aj = 0.16 from the dwarf comparison, 0.11 to $\sim$ 0 from the Pleiades one, and $\sim$ 0 from Alpha Per.  Assuming a spectral type of M6 or M7 moves the upper limit to \aj = 0.26.  We thus adopt an average \aj $\approx$ 0.13$\pm$0.1, similar to the F\&C01 result (\aj $\approx$ 0.13--0.26).   

Our derived extinction implies an intrinsic {\jks} of $\sim$ 1.1 in LS-RCrA~1\footnote{We use Rieke \& Lebofsky (1985) to correct the observed 2MASS $K_S$ for extinction.  The latter authors use the $K$ filter, not $K_S$; however, the small error incurred from our ignoring this difference in bandpass does not affect our results regarding NIR excess in LS-RCrA~1.}.  Even with no extinction, one gets \jks = 1.2.  For M6--M7 field dwarfs, {(\jk)}$_{CIT}$ $\sim$ 0.89--0.96 (Leggett 1992; i.e., {(\jks)}$_{2MASS}$ $\sim$0.92--1.00); for the same spectral types in the Pleiades, {(\jk)}$_{UKIRT}$ $\sim$ 1.0--1.25 (ZMR97; i.e., {(\jks)}$_{2MASS}$ $\sim$1.04--1.31).  For Alpha Per, {(\jk)}$_{2MASS}$ $\sim$0.94-1.22 (Barrado y Navascu\'es et al. 2001).  The largest possible \jks excess in LS-RCrA~1 is then $\sim$ 0.3 mag; the true value is probably significantly lower ($\sim$ 0.1, comparing our dereddened color to the average of the dwarfs, the Pleiades and Alpha Per).  This is within the errors expected from spectral type uncertainties, as the field dwarf, Pleiades and Alpha Per \jh ranges show.  Thus, even with our more accurate 2MASS photometry for LS-RCrA~1, we find no significant NIR excess suggestive of dust continuum emission, in agreement with F\&C01.   A disk may be more apparent in the L' band (e.g., in K-L'; Jayawardhana et al.  2003); this needs to be checked in future work.  The lack of NIR excess up to the K-band is not necessarily surprising; for example, most of the low mass accretors in IC 348 listed in Muzerolle et al. (2003; hereafter, MHCBH03) do not show any strong NIR excess (MHCBH03 quote $\Delta$$H-K$), while their optical veilings, and corresponding accretion rates, are similar to that in LS-RCrA~1 (see \S 3.4).  Since the peak of the SED in these cool, late -type objects is in the NIR, and their disks are also expected to be cooler and fainter than in higher-mass CTTs, it is hard to distinguish any disk emission from their intrinsic NIR photospheric flux.  Moreover, the magnitude of NIR excess also depends on the size of the inner disk hole.  Thus, the lack of an excess in LS-RCrA~1 is not terribly suggestive one way or another.  However, it is noteworthy that NIR excess also depends on disk inclination, rapidly decreasing as one moves from a face-on to edge-on geometry.  Now, as we discuss shortly, an edge-on orientation explains other, more puzzling characteristics of LS-RCrA~1; it is thus heartening that the lack if NIR excess is at least consistent with this geometry (though by itself not proof of it, given the other uncertainties in excess mentioned above).

\subsection{\lbol, \teff, Age and Mass}
As F\&C01 discuss, the distance to the R~CrA region is uncertain, with estimates ranging from 50 to 170 pc.  The value most commonly used is $\sim$ 130 pc (from Marraco \& Rydgren, 1981); moreover, Casey et al. (1998) have measured the distance to the double-line spectroscopic binary TY~CrA, deriving a value of 129$\pm$11 pc.  We therefore adopt 130 pc here, implying a distance modulus of 5.57.  A J-band bolometric correction (\bcj) is most appropriate for deriving the luminosity of accreting PMS objects (Luhman 1999).  Using the 2MASS $J$=15.18, our adopted values for \aj and distance, and a \bcj = 2.055 found by Leggett et al. (2002) for M6.5  dwarfs, we derive a bolometric luminosity \logten(\lbol/\lsun) = -2.73.  This is similar to the F\&C01 value of -2.63; the small difference (negligible compared to errors in \lbol arising from distance uncertainties) is due to slight variations in the observed $J$ magnitude and the bolometric correction.  

Luhman (1999) suggests a spectral type to \teff conversion scale for PMS M-types, which is intermediate between the \teff-scales for giants and dwarfs.  This scheme is specifically constructed to agree with the predictions of the theoretical HR diagrams of Baraffe et al. (1998; hereafter BCAH98) for young clusters; using it, F\&C01 derive \teff $\approx$ 2900$\pm$200K for LS-RCrA~1.  However, the theoretical tracks have significant uncertainties for very young low-mass objects, arising from inaccuracies in choice of initial conditions, treatment of convection and so on (Baraffe et al., 2002; Luhman et al., 2003); basing a \teff scale on their predictions for such objects may not be ideal.  It is therefore useful to have an independent method of estimating \teff for very low-mass PMS objects.  Mohanty et al. (2003; hereafter M03) suggest that this can be efficiently accomplished by comparing the profiles of the \teff-sensitive (and gravity {\it in}sensitive) TiO bandheads around 8440\AA~ to synthetic spectra, at high resolution (these are the same bandheads used for our spectral-type determination in Fig. 2).  This method is independent of the theoretical evolutionary model predictions; it also does not depend on a priori extinction estimates (over the relatively small wavelength regime analyzed, $\lesssim$ 100\AA, the extinction is constant, and is thus divided out during the relative normalization of the observed and synthetic spectra).  

We have carried out this analysis for LS-RCrA~1, using the latest synthetic spectra generated by Allard \& Hauschildt (Allard et al., 2001; Allard et al., in prep.).  Our results are displayed in Fig. 4; clearly, the 2700K model fits both LS-RCrA~1 and ChaH$\alpha$8 (which is the best PMS spectral type template for LS-RCrA~1, see Fig.2) very well.  Though we show models with a gravity of log \gv = 4.0 (appropriate for young low-mass objects, according to BCAH98), changing this value by $\pm$ 0.5 dex (which reasonably covers the range of possible PMS gravities for low masses) does not significantly change the derived \teff.  This analysis thus implies \teff $\approx$ 2700$\pm$100 K, somewhat lower than the F\&C01 value of 2900$\pm$200 K (in agreement with M03, who find the \teff implied by their spectral fits to mid- to late M PMS objects to be systematically lower than those in the Luhman scale by $\sim$ 100--200K; a similar downward revision, through improved spectral synthesis, has occurred recently in the dwarf \teff-scale at these spectral types (Leggett et al., 2000, 2001). We note that while dust opacity effects become important in late M-types, they do not appear for \teff $\gtrsim$ 2500K (Leggett et al., 2000; M03), so our lower \teff cannot be ascribed to an inadequate treatment of dust in the synthetic spectra.  We also point out that, if continuum veiling is affecting the observed depth of the TiO bands analysed here (our veiling results (\S 2.3) suggest the effect should be minimal at these long wavelengths), then the bands are intrinsically even deeper, implying an even lower \teff than we derive.  Of course, it is plausible that the synthetic spectra have systematic uncertainties that make our \teff determination inexact.  M03 analyse possible systematics in detail, and argue that they are likely to be quite small (of order no more than $\sim$100K); moreover, at least in early to mid-M field dwarfs, synthetic spectra appear quite accurate, with the \teff implied by them (Leggett et al. (2000)) being in good agreement (within $\pm$100K) with those recovered from recent empirical determinations based on interferometric radii measurements (Segransan et al., 2003; no empirical measurements exist yet for low-mass PMS objects).  We therefore suggest that our somewhat lower \teff of $\sim$ 2700K  (compared to the Luhman-scale value of $\sim$ 2900K, found by F\&C01) may be a more accurate value for LS-RCrA~1.

Determination of position on the H-R diagram, and thus mass and age, is somewhat more complicated.  In spite of any uncertainties in the theoretical evolutionary models, they are currently the only indicators of the evolutionary state of a young, very low-mass object (M03 present an alternative method of mass derivation independent of these tracks; however, this is beyond the scope of the present work, and also offers no independent measure of age).  We {\it must} therefore rely on some theoretical model or the other, to make further progress in our analysis.  We employ the BCAH98 models here; these are among the most widely used at present, and are the same ones used by F\&C01.  Here, the Luhman spectral type to \teff conversion scheme is indeed useful: though the absolute \teff implied by this scale (and the attendant mass and age, through comparison to the tracks) may be flawed as discussed above, the fact that it is specifically constructed to be compatible with the BCAH98 models enables a consistent {\it differential} analysis between various objects when comparing to these tracks.  On the other hand, assuming our independently derived \teff is more accurate in an absolute sense, it is certainly useful to check the mass and age predicted by the BCAH98 models when this \teff is adopted.  It is thus most useful to perform the analysis using both \teff estimates, under the assumption that this brackets the real evolutionary situation. 
 
Figure 5a displays the theoretical H-R diagram and evolutionary tracks of BCAH98, and the location of LS-RCrA~1 implied by our \lbol estimates, combined with either our \teff or that implied by Luhman for spectral type M6.5.  We see that the object appears to have a mass $\sim$ 0.04$\pm$0.01 \msun with our \teff, and $\sim$ 0.08 \msun with the Luhman value.  Similar results are achieved with other tracks (Burrows et al. 1997; D'Antona \& Mazzitelli 1994; Siess et al. 2000).  Thus, regardless of which of the two \teff estimates one prefers, it appears fairly certain that this object is  at or below the substellar boundary.  
Note that a 1-10 Myr isochrones  and \teff=2700 K would produce a mass in the range 0.035-0.030 \msun (0.072-0.065 \msun in the case of \teff=2900 K).

From Fig. 5a, we also infer an age of $\sim$ 20 Myr (and larger than 7 Myr), with our \teff estimate.  Using the Luhman \teff implies an even larger age of $\sim$ 50 Myr.  Now, the assumed age for the R~CrA cloud is $\sim$3 Myr (Wilking et al. 1997).  N00 have found ages ranging from 1 to 10 Myr for their ROSAT-selected R~CrA targets; the 2 accreting classical T Tauri stars in their sample seem $\sim$ 10 Myr old.  Fig. 5a shows that our measurements for LS-RCrA~1 are only marginally compatible with 10 Myr.  N00  used the Luhman \teff scale for the M-types in their sample; thus their stated values are most profitably compared to the 50 Myr age for LS-RCrA1 derived using the same scale.  While PMS \teff have not yet been derived through spectral synthesis at the M4 and earlier types that make up most of the N00 sample, the general trend in the mid- to late M's, as we have noted above, is that the latest synthetic spectra imply \teff somewhat lower than the Luhman scale.  Assuming this trend continues to earlier types, the \teff derived from spectral synthesis would yield an even younger age for the N00 sample than these authors derive.  This is certainly true at least for the later M types in their sample (for instance, their M6 object, with estimated age $\sim$ 2.5 Myr on the Luhman scale, would be $\sim$ 1 Myr old if its \teff were $\sim$200K lower, as we suggest).  What this means is that, the particular \teff-scale one chooses does not mitigate the apparent age {\it difference} between this object and the rest of the known R~CrA sample.  Using either the Luhman scale or ours merely shifts (all the) objects along the \teff axis; the choice of \teff scale does not change the fact, as F\&C01 pointed out, that LS-RCrA~1 appears much fainter than one would expect for a cluster member, and hence relatively older.  

However, our spectra show that LS-RCrA~1 is a bona-fide PMS object, while our \vrad measurements are consistent with its being a true member of the R~CrA complex.  (\S3.1).  Moreover, it also shows signs of ongoing accretion (\S 3.4).  Taken together, these facts lead one to conclude that either (1) LS-RCrA~1 is a member of R~CrA and star-formation has continued in this region over tens of millions of years, {\it and} that LS-RCrA~1 was formed earlier than all the known stars in the complex, {\it and} that it continues to accrete robustly after 20 Myr, or (2) the apparent old age of this object is artificial\footnote{A third possibility is that our slight offset in \vrad from the N00 values for other CrA members, while not significant given our measurement uncertainty and the small NOO sample size, is actually real and a sign of non-membership.  In that case, either LS-RCrA~1 is much younger, and appears underluminous because it is much further away, or it is at $\lesssim$ 130 pc (our assumed distance to the R~CrA complex), and continues to accrete at an age $\gtrsim$ 20 Myr.  Accretion at the latter ages seems highly unlikely.  Alternatively, for a more plausible accretion age limit of $\lesssim$ 10 Myr, and no true underluminosity, the object would have to be $\sim$ 200--300 pc away.  This too seems very implausible, given the unlikelihood of seeing a background star-forming region sitting just along the line of sight to a nearby one, and moreover given that the CrA region is a dense molecular region with high background extinctions, largely precluding the visibility of background sources.}.    

The first is improbable; we thus concentrate on the second possibility. N00 do not consider even the 10 Myr age spread they find to be real, and ascribe it instead to errors in derived luminosity arising from a spread in distances to their targets.  Nevertheless, using the same mean distance as we do (130 pc) leads N00 to infer a maximum age of 10 Myr over their entire sample;  if LS-RCrA~1 appears sub-luminous, and hence older, because it is actually further away than we assume, then its distance must be much larger than to all the other R~CrA members.  This too does not seem plausible.  The final option is that a real physical effect makes LS-RCrA~1 appear older/fainter. 
This option is supported by the location of the object in a diagram displaying the surface gravity (logg=4.0 in the
case of LS~RCrA-1,a s derived from the spectrum) against the effective temperature (Figure 5b). Both quantities
are intrinsic to the object, and are unaffected by errors in the distance, interstellar reddening, and
bolometric correction. This diagram indicates that its age, about 8 Myr, is within  the estimated value for
the association (in any event, it should be younger than 20 Myr) 
and that the mass is in the range 0.025-0.045 M$_\odot$ (or slightly larger than 0.070  M$_\odot$ 
if we take the \teff derived by F\&C00).

 As F\&C01 note, the fact that it is accreting immediately suggests two possibilities.  One is that the process of accretion changes the luminosity and \teff of the object; the other is that an edge-on disk blocks much of its light, making it appear fainter.  

The first scenario has been theoretically explored by various authors for higher mass stars (see F\&C01 and references therein).  Accretion is found to increase both the luminosity and \teff of the star.  The net result is to make it seems older  than it really is, when compared to theoretical tracks that do not take accretion into account; the effect becomes stronger with increasing accretion rate.  A 1 \msun star accreting at 10$^{-6}$ {\msun}yr$^{-1}$ can seem older by more than a factor of 2, while a rate of $\lesssim$ 10$^{-8}$ {\msun}yr$^{-1}$ has negligible effect on the inferred age (Hartmann, Cassen \& Kenyon, 1997).  Indeed, this effect might explain the otherwise slightly puzzling fact that the two (higher mass) CTTs in the N00 sample appear to be among the oldest members of the sample.  Can such a procees explain the position of LS-RCrA~1 on the H-R diagram?  As discussed in \S 3.4, an accretion rate greater than $\sim$ 10$^{-9}$ {\msun}yr$^{-1}$ does not appear viable for this object.  However, one might posit that, in very low mass objects, the age effect described above may still be apparent at lower accretion rates than it is in higher mass stars.  Theoretical calculations to test this idea are beyond the scope of this paper.  However, we attempt to test it empirically, by comparing the inferred ages of very low mass accretors and non-accretors in IC 348.  Using the BCAH98 tracks, Luhman (1999) found a mean age ($\sim$ 3 Myr) and age range ($\sim$ 1--10 Myr) for this cluster identical to that derived for R~CrA by N00; moreover, a number of low-mass accretors have been identified in this cluster (Jayawardhana, Mohanty \& Basri 2003, hereafter JMB03; MHCBH03).  The combination of these facts makes IC 348 ideal for comparison to R~CrA.  

IC-348 165, 205, 382, 415, 173 and 336 are all known accretors (JMB03; MHCBH03).  Their spectral type range is $\sim$ M5--M6.5 (Luhman 1999), roughly similar to that estimated for LS-RCrA~1.  Accretion rates for all have been derived by MHCBH to lie between $\sim$ 10$^{-9}$ and 10$^{-11}$ {\msun}yr$^{-1}$; i.e, from about the same rate we infer for LS-RCrA~1 to 1--2 orders of magnitude lower.  The magnitudes of their observed veilings and 10\% \hal widths straddle those seen in LS-RCrA~1. MHCBH03 find that two of them (IC-348 165 (M5.25) and 173 (M5.75)) exhibit forbidden OI emission ($\lambda\lambda$6300, 6363) indicative of outflows, as LS-RCrA~1 does.  In short, any accretion-related effects in this sample of IC-348 objects should, on the whole, be roughly similar to those in LS-RCrA~1.  

However, a perusal of their positions on the H-R diagram, as depicted by Luhman (1999), shows that all of them lie comfortably within the 1--10 Myr age range covered by the known {\it non}-accreting low-mass members of IC 348; none of them appears to be much older than expected, in contrast with LS-RCrA~1, nor is there any perceptible trend towards larger inferred age with increasing accretion rate.  These results are independent of any systematic shifts in the Luhman \teff scale that might be suggested by spectral synthesis (discussed earlier); such shifts would not affect the relative ages inferred.  We conclude (1) that accretion rates $\lesssim$ 10$^{-9}$ {\msun}yr$^{-1}$ do not seem to give rise to age discrepancies in very low-mass accretors, and (2) that the anomalous position of LS-RCrA~1 on the H-R diagram is unlikely to arise from accretion-related effects. 
 Again, this is reinforzed by the age and mass estimate derived from the location of the object in Figure 5b..

Given these results (based ultimately on accretion rates derived from our optical veiling measurements), we are led to favor the disk scenario.  Specifically, we suggest that a nearly edge-on disk occults the central object in LS-RCrA~1.  The latter is seen only in light scattered off the disk surfaces, reducing its apparent luminosity far below the intrinsic value.  The upshot is that LS-RCrA~1, a true member of the R~CrA complex, appears artificially sub-luminous and old; correcting for the reduction in photospheric flux by the intervening disk would remove the seeming age discrepancy between it and the other R~CrA members.  The validity of this claim will have to be checked through further observations, e.g., deep high angular-resolution imaging.

Recently, Comer\'on et al. (2003) have argued against the edge-on disk scenario, and in favor of an accretion-induced shift on the H-R diagram (which we argue against, above), to explain the observed sub-luminosity of LS-RCrA~1 (and another similar object in Lupus, Par-Lup3-4).  Their  suggestion is based on an alleged reduction in flux of \hal and other accretion signatures in the presence of an edge-on disk.  We discuss, and rebut, their claim at the end of the next section.

\subsection{Accretion and Mass Outflow}
Figure 3 displays the average low and medium resolution spectra (top and middle panels). Several intense forbidden lines, as well as H$\alpha$ and the CaII infrared triplet have been labeled.  The measured  equivalent widths (W) for each night are listed in Table 2.  

The bottom panels in Figure 3 also show the H$\alpha$ profile (with the instrumental profile at medium resolution as a dotted line) and the spectral range around LiI6708 \AA{} and HeI6678 \AA{}.  At this resolution, the we cannot distinguish any asymmetry in the H$\alpha$ profile, but certainly the H$\alpha$ is much broader than the instrumental profile (120 km/s at FWHM).  Inspection of individual low- and medium-resolution  spectra suggests that some variability might be present in the H$\alpha$ equivalent width (Table 3).

Our echelle high resolution spectra, obtained 2 months later, show the broad, somewhat asymmetric H$\alpha$ line (Fig. 6a), with a width of 316 \kms at 10\% of maximum intensity, well above the $\sim$ 200 \kms threshold for accreting substellar objects proposed by Jayawardhana, Mohanty \& Basri (2003). F\&C01 had indeed concluded that the \hal is mainly produced in an accretion flow, and not in a shock produced by a jet/outflow, by examining the ratio of its equivalent width to that of the forbidden [SII] emission from low-resolution spectra (W$_{H\alpha}$ / (W$_{[SII]6717}$ + W$_{[SII]6731}$)~).  However, the equivalent width of \hal (and to a lesser extent, the forbidden lines) tends to be significantly overestimated at low- compared to high-resolution; for instance, we find 53\AA~ at high-resolution and 221\AA~ at low (and F\&C01 measured  360\AA~).  The ratio of the \hal to [SII] line widths at high-resolution is then only 6.6: still somewhat higher than the maximum value of 4 seen in HH objects (F\&C01 and references therein), but significantly less than the 11 derived by F\&C01 (or the 10 implied by our low-resolution data).  Thus, in LS-RCrA~1, this line ratio does not appear to be as strong an indicator of the \hal being associated with accretion, as opposed to with an HH shock, as F\&C01 had proposed from their low-resolution analysis.  This is important, since there is a small possibility (discussed shortly) that the {\it forbidden} line emission in LS-RCrA~1 may arise in an unassociated HH knot in the line of sight; it is necessary to rule this possibility out at least for the \hal line.  The \hal emission profiles in the high-resolution spectra enable this: the broad wings of the line unambiguously point to its genesis in an accretion flow (e.g., Muzerolle, Calvet \& Hartmann, 1998a).  In other words, whatever the origin of the forbidden line emission, LS-RCrA~1 is definitely an accretor.  This conclusion is also supported by the presence of optical veiling in our spectra (\S 2.3), which is due to continuum excess emission generated in an accretion shock.  Finally, it is corroborated by the \ca IR triplet (Fig. 6a) and permitted OI (8446\AA; Fig. 2) emission lines, which exhibit the broad profiles (FWHM $>$ 100\kms) expected from accretion, but not from chromospheric activity or production in an outflow-related shock.
 
Our veiling estimates also allow us to derive an approximate accretion rate for this object.  MHCBH03 give theoretical predictions for the accretion rate as a function of wavelength-dependent veiling and spectral type.  Comparing our values to their predicted ones, for a spectral type of M6, yields an accretion rate of 10$^{-10}$ $\lesssim$ \.{M} $\lesssim$ 10$^{-9}$ {\msun}yr$^{-1}$.  For M6.5 (not explicitly shown in MHCBH03), the precise rate inferred would be slightly less (since the same excess emission would give a larger veiling compared to the fainter photosphere of a later type), but still within the range specified above.  Our inferred rate is supported by the fact that MHCBH03 derive a very similar value, \.{M} = 10$^{-9.3}$ {\msun}yr$^{-1}$, for IC-348 415, which too has a spectral type of M6.5 (Luhman et al., 2003) and veiling estimates (MHCBH03) similar to ours for LS-RCrA~1.  

We further note that an accretion rate significantly larger than 10$^{-9}$ {\msun}yr$^{-1}$ appears highly unlikely for LS-RCrA~1.  GM Tau is a PMS object in the Taurus Auriga complex that has long been identified as a continuum T Tauri star; i.e., its accretion-generated optical veiling completely masks the intrinsic photospheric spectrum at blue optical wavelengths.  Recently, WB03 have been able to derive a spectral type of M6.5 for this object, from the reddest regions of high-resolution spectra where the veiling is least.  They go on to derive veilings of at least 1 (ie, excess emission comparable to photospheric), even at comparatively red wavelengths of 8000--8400 \AA, and of upto 2 in bluer regions.  However, even with such high veilings, WB03 infer an accretion rate of $\sim$ 10$^{-8.5}$ {\msun}yr$^{-1}$ for GM Tau.  This is comparable to the lower end of values commonly seen in higher-mass, {\it non}-continuum classical T Tauri stars.  In other words, a reasonably small accretion rate is enough to dominate, in the optical, over the faint photosphere of an M6.5 object such as GM Tau (but not that of brighter, higher mass T Tauri stars).  Since LS-RCrA~1 also has a spectral type $\sim$ M6.5, but is clearly not a continuum star - it has easily distinguishable photospheric absorption features in our optical spectra, which lead us to derive veilings vastly less than in GM Tau - we conclude that its accretion rate must be significantly lower than that of GM Tau.  Given the WB03 estimate for the latter, \.{M} $\lesssim$ 10$^{-9}$ {\msun}yr$^{-1}$ appears quite appropriate for LS-RCrA~1.  

At this juncture, it is worth noting that the 10\% \hal width of $\sim$ 316 \kms we find for LS-RCrA~1, in our high-resolution data, is substantially higher than the $\sim$ 213 \kms quoted by JMB03 for IC 348-415 from similar resolution spectra, and higher too than the $\sim$ 272 \kms in MHCBH03 from medium-resolution spectra (even though reduced resolution tends to overestimate the \hal width).  This is in spite of the fact that the two objects have very similar spectral type, veiling and inferred accretion rates, which should make the physical conditions in the accretion flow, and in particular the infall velocities, quite similar (any differences in mass and radius, despite the similar spectral type, should not affect the infall velocity much, as MHCBH03 discuss).  The simplest way to explain the larger \hal width is to posit that LS-RCrA~1 is seen edge-on (i.e., its rotation axis is perpendicular to the line of sight).  In this case, we see more high-velocity infalling material than in the pole-on situation, resulting in broader \hal line-wings (MHCBH03; Muzerolle, Calvet \& Hartmann 2001).  From fitting its \hal profile, MHCBH03 derive an inclination angle (measured from the line of sight) of 45$^{\circ}$ for IC 348-415; our argument here implies that the angle should be much higher for LS-RCrA~1, making it closer to edge-on.  Indeed, for several IC-348 objects which have spectral types and 10\% \hal widths similar to LS-RCrA~1, and accretion rates in the same range (e.g., IC-348 205: M6, high-resolution 10\% width $\sim$ 338 \kms (JMB03), \.{M} $\sim$ 10$^{-10}$ {\msun}yr$^{-1}$ (MHCBH03)), MHCBH03 do in fact derive nearly edge-on orientations (e.g., 80$^{\circ}$ for IC 348-205).  Thus, the relatively large \hal 10\% width in LS-RCrA~1 provides independent evidence of the edge-on scenario proposed to explain its sub-luminosity.   

Finally, we point out that Comer\'on et al. (2003) have derived an accretion rate for LS-RCrA~1 using the strength of the CaII IRT emission (specifically, the equivalent width of the 8542\AA~ component) in their spectra.  Employing the relation between CaII flux and accretion rate given by Muzerolle, Hartmann \& Calvet (1998b), they find \.{M} $\sim$ 2.8$\times$10$^{-10}$ {\msun}yr$^{-1}$.  In our high-resolution spectra, the equivalent width of the 8542\AA~ line is about 2.5 times larger than in Comer\'on et al.'s data (9\AA~ versus 3.6\AA); using the same formalism and CaII component, therefore, we find an accretion rate of $\sim$ 7$\times$10$^{-10}$ {\msun}yr$^{-1}$.  The difference in our CaII widths may be due to emission variability.  However, it is also probable that our high-resolution data are more accurate: Fig. 6a shows that the CaII lines are very broad (full width at 10\% of peak intensity $\sim$ 200 \kms), but with relatively low peak intensity; the equivalent widths of such lines would be underestimated in lower-resolution spectra.  The interesting point, however, is that both our and Comer\'on et al's values are consistent with the accretion rates implied by our veiling measurements (10$^{-9}$--10$^{-10}$ {\msun}yr$^{-1}$).  This agreement needs to be understood in the context of the edge-on scenario, as we now discuss.  

The accretion rate derived from the veiling measurements should be largely independent of any occlusion due to an edge-on disk, since presumably the flux from the photosphere and the veiling shock emission, which arises on the stellar surface, are reduced by occlusion by similar fractions.  Occlusion, however, may affect the rate derived from the CaII emission, depending on where the line is formed.  If the emission arises mainly from the shock region on the stellar surface, then the accretion rate inferred from it would be an underestimation:  while the equivalent width would be largely unaffected (since it is a measure of the emission strength relative to the underlying continuum, and both would be similarly reduced by occlusion, as in the veiling case), the {\it absolute} flux of the CaII emission (and hence derived accretion rate) would be underestimated (since this flux, in Comeron et al. and our analogous calculation, is derived by scaling the equivalent width by the observed photospheric flux, which is lower than the intrinsic photospheric flux due to occlusion).  On the other hand, if the CaII IRT emission arises mainly in the magnetospheric infall region, and thus remains unoccluded (as we have argued for the \hal magnetospheric line), then the inferred accretion rate may be expected to be approximately correct: while the observed equivalent width of the CaII lines would be larger than in the absence of occlusion (since the underlying photosperic flux is now reduced, but not the line emission), the absolute emission flux is derived by scaling this enhanced width by the observed photospheric flux, which is {\it reduced} by the same fraction; the derived absolute flux of CaII emission would thus be roughly accurate.   

The above arguments suggest that the agreement between the accretion rates derived from the CaII emission and the veiling measurements are explicable in an edge-on scenario only if the CaII IRT we observe arises in the magnetospheric infall region, and remains largely unobstructed by the disk.  In this case, one might expect the CaII line profiles to be characterized predominantly by a broad component (BC).  As Fig. 6a reveals (and mentioned earlier), this is indeed the case: the observed high-resolution profiles are broad in all three IRT lines, and we see no indication of a narrow component (NC) that would be expected from emission from the shock region (this is not to say that a NC does not exist; it may well be that occlusion by a disk has reduced its intensity to below the levels in the BC).  In this context, we add that to the best of our knowledge, only the CaII BC, and not the NC, has been shown to accurately trace accretion rates.  For instance, in the vast majority of stars used to demonstrate the relationship between accretion rate and CaII flux in Muzerolle, Hartmann \&  Calvet (1998b), the BC widths greatly dominate over the NC ones\footnote{Only 3 stars in their sample have a dominant (but small) NC.  Moreover, in all 3 cases, a BC is not observed at all and the accretion rates are also the lowest.   It is thus unclear, even in these 3 cases, whether it is the small NC or the even smaller (unobserved) BC that better signifies the very low accretion rate.}.  It is thus perhaps unsurprising that considering the flux contribution of the BC alone, and not the unobserved (but not necessarily insignificant) NC, yields the correct accretion rate (assuming our rates from veiling are valid).  To summarize: the agreement between the veiling and CaII accretion rates is compatible with an edge-on disk only if most of the line emission arises in the infall region and is not occluded by the disk; this implies that the line-profiles should be broad; this prediction is in agreement with our observed high-resolution profiles.  We opine, therefore, that our CaII observations are consistent with an edge-on disk scenario.  

Fig. 6b illustrates several much narrower profiles belonging to forbidden lines of Oxygen, Sulfur and Nitrogen, present in our high-resolution spectra.  As noted earlier, and discussed by F\&C01, these lines are suggestive of a jet/outflow associated with LS-RCrA~1.  However, it is noteworthy that while the accretion rate we infer for this object is consistent with the rates derived in many other very low-mass accretors of similar spectral type (MHCBH03), none of the latter has been reported to exhibit such a plethora of forbidden lines (though a couple in IC 348, as mentioned earlier, show forbidden [OI] suggestive of outflows).  In other words, any jets or outflows that might exist in such objects appear in most cases to be too weak to generate such strong lines, even though the accretion rates are similar to that of LS-RCrA~1, and outflow rates are expected to be directly related to accretion ones.  Two possible explanations may be suggested.

One is that the forbidden lines are not actually associated with an outflow from LS-RCrA~1 itself, but arise instead in an HH knot that is driven by some other source and happens to lie in the line of sight. There are indeed a number of such knots close to LS-RCrA~1.  To its immediate east, and extending roughly along a NE-SW axis, are 7 knots (HH 96, 97, 98, 99, 100, 101 and 104); the closest of these (HH 96) is $\sim$ 1.5' away from our object.  HH 100 IRS is thought to be the driving source for HH 96--101, while HH 104 appears to be associated with the Herbig Ae/Be star R~CrA.  To the north-west of LS-RCrA~1, $\sim$ 4.7' away, is another HH knot, HH 82, which is thought to be driven by the source S RCrA.  Given the number of energy sources and HH knots in the vicinity of LS-RCrA~1, it is possible that there is another (as yet unidentified) HH knot, not physically associated with LS-RCrA~1, that intersects our line of sight to it.  Whether this is indeed the case needs to be investigated through further observations.  However, this scenario appears improbable a priori.  Given that our high-resolution data were taken with a 5'' long by 1'' wide slit, an intervening HH knot would have to lie almost exactly on LS-RCrA~1 to superimpose its lines on our spectra.  Furthermore, the positions of the observed lines are commensurate with zero velocity in the rest-frame of LS-RCrA~1, which seems unlikely for a random superimposed HH shock.  Finally, our low- and medium-resolution spectra do not show  any indication of spatially extended emission in the 2-D spectra. Taken together, these facts make the line-of sight HH knot scenario seem quite implausible.  

The other possibility is that the forbidden lines are indeed associated with a jet/ouflow from LS-RCrA~1, but enhanced compared to other similar accretors due to the presence of an edge-on disk.  As discussed in \S 3.3, such a disk would greatly reduce the photospheric flux from the central object; however, it would leave the outflow emission, which arises above and below the plane of the disk, relatively unaffected.  The net result would be an enhancement of the forbidden lines relative to the photosphere, compared to other sytems not seen edge-on, in which the light from the photosphere overwhelms the forbidden line emission.  This is the most plausible scenario, and and is yet another line of evidence suggesting that LS-RCrA~1 is an edge-on system.  

Furthermore, none of our forbidden lines show any high-velocity components (HVCs) in the high-resolution data (Fig. 6a), which suggests that the outflow rate is small, consistent with the low accretion rate we find\footnote{There is an emission feature (6297.9\AA) close to the [OI]6300 line that looks like an HVC.  However, this is an insufficiently subtracted sky line, that appears in both our extracted sky spectrum and in our spectrum of Cha\hal 8 (a non-accretor), and is also mentioned in the list of skylines by Hanuschik (2003).  The relative shortness of the slits on Magellan MIKE sometimes make accurate sky-extraction and subtraction difficult for faint objects; this is the case here.}. We note that Comeron et al. (2003) derive an outflow rate that is more than an order of magnitude higher than our accretion rates; however, their outflow rates are based on the assumption that the entire forbidden line emission consists of an HVC, and they caution that their values are serious overestimations if the HVC is small or non-existent, as we find.  In fact, the absence of HVCs in our high-resolution spectra can be used to put an upper limit on the outflow rates.  Our detection limit for line emission is $\sim$ 0.3\AA.  Using this value as an upper limit for HVC strength, and combining it with the same equations (and thus underlying assumptions) employed by Comeron et al., we find an outflow upper limit of $\sim$ 4$\times$10$^{-11}$ {\msun}yr$^{-1}$ from the absence of an HVC in the [OI]6300 line, and $\sim$ 6.5$\times$10$^{-11}$ {\msun}yr$^{-1}$ from the lack of an HVC in the [SII]6731 line. Given the simplistic assumptions here, these low outflow rates are quite consistent with our (slightly higher) accretion rates.  Moreover, in the context of the X-wind model, the outflow rate is expected to be roughly half the infall one; the accretion rates implied by the above outflow estimates are then $\sim$ 10$^{-10}$ {\msun}yr$^{-1}$, in keeping with our accretion estimates through other means.  We have argued above that the prominence of the forbidden lines in LS-RCrA~1 is probably not due to any extraordinary strength of its outflow, based on its low accretion rate and the proportionality between accretion and outflow rates established for other CTTs; the lack of HVCs in these lines suports this conclusion.  Moreover, we see no evidence for strong asymmetries in the forbidden lines (Fig 6a; a slight asymmetry may be present in the OI lines).  This is precisely as expected for a disk seen close to edge-on, so that the jet/outflow is perpendicular to the line of sight.  In summary, the lack of both HVCs and asymmetry in the forbidden lines suggests a low accretion rate combined with an edge-on geometry.

Finally, we note that if the edge-on scenario is correct, so that jet is in the plane of the sky, its velocity component in the radial (line-of-sight) direction may be small.  In this case, any HVC would be swamped by the low-velocity component, and the outflow rate may be somewhat higher than our rough estimate above.  However, since this only happens if the disk is edge-on in the first place, it is not an argument against the edge-on hypothesis.  

As mentioned earlier, Comer\'on et al. (2003) contend that an edge-on disk is not viable.  The essence of their argument is that such a disk would also occult accretion signatures such as \hal, that arise close to the stellar surface; as a result, the intense emission actually observed in these lines in LS-RCrA~1 would not be expected.  There are two simple answers to this: (1) intense accretion signatures {\it have} been observed in other edge-on systems, and (2) the generally accepted geometry of magnetospheric accretion is consistent with these observations.  Regarding the first point, as Luhman (2003) points out, intense \hal is seen in the edge-on system MBM 12A 3C (Jayawardhana et al. 2002), as well as in the PMS object KH15D during its periodic occultations by circumstellar material.  In the latter case, the \hal flux remains roughly constant while the underlying photospheric flux drops drastically during an occultation (Hamilton et al. 2003).  From a theoretical point of view, it is thought that \hal in accreting T Tauri objects arises in the infalling accretion column, and not just in the accretion shock on the stellar surface (e.g., Muzerolle, Calvet \& Hartmann 1998a).  In the usually adopted geometry, these magneto-centrifugally channeled accretion columns rise above the plane of the disk before finally landing on the stellar surface.  It is quite plausible therefore that the infall regions producing \hal rise sufficiently above the disk to not be fully occulted (analogous to the proposed situation in the forbidden emission lines, above).  This is the most natural solution, for example, for the KH15D observations.  The same argument can also be made for other lines arising in the magnetospheric infall region, e.g., the broad components of the CaII IRT lines; CaII is indeed observed in the edge-on system MBM 12A 3C\footnote{no observations of KH15D in the CaII region, during eclipse, have yet been presented in the literature.}.  We see no reason, therefore, to rule against the presence of an edge-on disk in LS-RCrA~1.  


\section{Summary and Conclusions}
We have collected low-, medium- and high-resolution spectra of the low mass object LS-RCrA~1 and extracted the 2MASS infrared photometry of the All Sky release.  Previously, only low-resolution spectra and photometry with significant error bars were available.  With this new wealth of data, we have been able to confirm the PMS status of this object and its membership in the R~CrA star-forming region.  Our \teff, luminosity and surface gravity estimates, combined with theoretical evolutionary tracks, imply a mass $\sim$ 0.035 \msun and an age of 8 Myr, well into the substellar regime.  We have also confirmed ongoing accretion in this object, from various line profiles and the presence of optical veiling, and derived a mass accretion rate of $\sim$ 10$^{-9}$--10$^{-10}$ {\msun}yr$^{-1}$ from the veiling.  We find that LS-RCrA~1 indeed appears to be sub-luminous, or older, compared to what is expected for an R~CrA member.  We do not believe this effect is due to accretion-related phenomena that perturb this objects's position on the H-R diagram: a large sample of other low-mass accretors with similar accretion rates do not exhibit any such behaviour.  We suggest instead that this effect is caused by the presence of an edge-on disk.  This interpretation is supported by the presence of unusually intense forbidden line outflow signatures combined with a low accretion rates and lack of HVCs in these lines (signifying either correspondingly low outflow rates, which suggests an edge-on disk when combined with the prominence of the outflow emission lines; or that the radial component of the jet velocity is small, direct evidence for a jet aligned in the plane of the sky and thus an edge-on disk).  An edge-on orientation is also consistent with the relatively broad wings observed in the \hal emission, the lack of assymetries in the forbidden lines and the lack of any significant NIR excess.  We contend that the ease with which an edge-on disk can simultaneously explain all these {\it independent} puzzling aspects of LS-RCrA~1 makes its presence around it an appealing explanation.  We point out that such a disk does not necessarily preclude intense accretion signatures, contrary to a recent suggestion:  the coexistence of the two might be expected on theoretical grounds (i.e., accretion columns that rise above the plane of the disk) and is also observed in at least two sources.  High-angular-resolution imaging may be able to confirm the presence of an edge-on disk and a jet/outflow in this system.  Our recent work has shown that a significant fraction of young sub-stellar objects harbor accretion disks. Our present results for LS-RCrA~1, together with those of F\&C01, strongly suggest that such objects can also drive mass outflows, strengthening the analogy with higher mass classical T Tauri stars.  This lends further credence to the idea that brown dwarfs and low-mass stars share a common formation mechanism.

\acknowledgements
We thank Diane Paulson for assistance with the observations and the
Magellan staff for outstanding support. We are grateful to Kevin Luhman
and Jochen Eisl\"offel for useful discussions. We are especially indebted to the referee, F. Comer\'on, for invaluable suggestions and comments.  This work was supported 
in part by NSF grant AST-0205130 to R.J.; DByN is indebted to the Spanish
``Programa Ram\'on y Cajal'' and AYA2001-1124-C02; S.M. gratefully 
acknowledges the support from the SIM-YSO group. This publication makes 
use of data products from the Two Micron All Sky Survey.


\newpage

\begin{table}
\caption{Properties of LS-RCrA~1.}
\begin{tabular}{lr}
\hline
   Time       &  W(H$\alpha$) \\
    (days)    &   (\AA)       \\

\hline
\hline
Distance    (pc)          & 130$^1$                  \\
age         (Myr)         &  20/8$^2$           \\
Mass        (M$_\odot$)   &(0.040/0.035)$^2$$\pm$0.010 \\
E(B-V)                    & 0.15$\pm$0.09   \\
A$_J$                     & 0.13         \\
Teff        (K)           & 2700$\pm$100  \\
Sp.Type                   &  M6.5$\pm$0.5 IV    \\
Log Lum     (L$_\odot$)   & -2.73/-2.64$^3$    \\
J (2MASS)                 & 15.178$\pm$0.052    \\
H (2MASS)                 & 14.526$\pm$0.061    \\
Ks (2MASS)                & 13.972$\pm$0.053    \\
r(6200 \AA)               &  1.00$\pm$0.10    \\
r(6750 \AA)               &  0.25$\pm$0.05    \\
r(7150 \AA)               &  0.15$\pm$0.05    \\
FWHM(H$\alpha$)   (km/s)  & 316 \\
$v~sini$          (km/s)  &  18$\pm$5   \\
RV                (km/s)  &  2$\pm$3   \\
W(LiI6708)        (\AA)   &  0.74$\pm$0.14 \\
Log Lum(H$\alpha$)/Lum(bol) & -2.68/-3.30$^4$ \\
\hline
\end{tabular}
\\
$^1$ Other values range from 70 to 150 pc.\\
$^2$ As derived  from \teff and Luminosity and \teff and surface gravity, respectively.\\
$^3$ Derived from $J$ and $K_S$, respectively.\\
$^4$ Log L(H$\alpha$)/L(bol)=-2.47 using W(H$\alpha$)=260 \AA{}
measured by F\&C01 in a low resolution spectrum.\\
\end{table}

\begin{table*}
\caption{Equivalent widths in \AA. Note the different spectral resolution 
for each night. 
}
\begin{tabular}{lrrrrr}
\hline
                  & March 9       & March 10      & March 11      & May 9    & F\&C01 \\
                  & R=620         & R=2,600       & R=2,600       & R=19,000 & R=500  \\
\hline                                                               
\hline                                                               
  $H\epsilon$ 3970&  out of range & out of range  &  out of range &  7.6 2.9 &   --  \\
  $[SII]$ 4069    &  out of range & out of range  &  out of range &  3.9 1.4 &   --  \\
  $H\delta$ 4102  &  out of range & out of range  &  out of range &  6.6 3.9 &   --  \\
  $H\gamma$ 4341  &  out of range & out of range  &  out of range &  6.6 3.9 &   --  \\
  $H\beta$  4861  &  out of range & out of range  &  out of range & --   --  &   --  \\
  $[OI]$ 5577     &   12.86  3.13 & out of range  &  out of range & --   --  &   --  \\
  $[OI]$ 6300     &   51.85  4.67 &   59.37  7.64 &   64.35  8.93 & 13.9 2.0 &  92   \\
  $[OI]$ 6364     &   13.35  1.52 &   16.73  2.28 &   19.16  2.74 &  4.8 0.5 &  25   \\
  $[NII]$ 6548    &    --     --  &    1.26  9.34 &    0.82  0.24 & --   --  &  --   \\
  $[NII]$ 6581    &    --     --  &    4.96  0.79 &    3.78  0.58 &  1.8 0.2 &  15   \\ 
  $H\alpha$ 6563  &  221.17 29.20 &  152.67 21.57 &  163.10 14.25 & 53.0 3.6 & 360   \\
  $HeI$ 6678      &    2.13  0.31 &    2.05  0.43 &    2.50  1.00 &$\le$0.3  &   1.5 \\ 
  $[SII]$ 6717    &    7.81  2.08 &    8.70  0.74 &    7.00  0.92 &  2.3 0.3 &  12.4 \\ 
  $[SII]$ 6731    &   14.15  2.40 &   15.64  1.84 &   13.88  1.28 &  5.7 1.1 &  19.2 \\ 
  $[OII]$ 7319    &6.01  0.76$^1$ &    2.23  0.41 &    2.21  0.24 &  1.3 0.2 &   3.6 \\ 
  $[OII]$ 7329    &    --     --  &    1.69  0.22 &    1.95  0.15 &  1.5 0.1 &   3.6 \\ 
  $CaII$ 8498     &   11.00  1.12 & out of range  & out of range  & 11.5 2.1 &   4.0 \\
  $CaII$ 8542     &   10.24  1.25 & out of range  & out of range  &  9.0 0.8 &   3.6 \\
  $CaII$ 8662     &    7.13  1.43 & out of range  & out of range  &  9.6 1.2 &   2.5 \\
\hline
\end{tabular}
$\,$\\
$^1$ blended with [OII] 7330\\
%
%
\end{table*}

\begin{table}
\caption{Note that these equivalent  widths were measured with the 300 l/mm
during the first night (March 9th), whereas we used the 1200 l/mm grating the
 next two (March 10th and 11th).
The MIKE echelle spectrograph was used during the fourth night, two months later (May 9th). 
Time in JD-2452700.0. Errors can be estimated as 15 \% in the first case and 10\% for the
other datasets.}
\begin{tabular}{lr}
\hline
   Time       &  W(H$\alpha$) \\
    (days)    &   (\AA)       \\
\hline
\hline
708.83143 &  214.7  \\ 
708.83884 &  242.5  \\ 
708.84626 &  235.0  \\ 
708.85373 &  189.0  \\ 
709.85140 &  168.8  \\ 
709.86578 &  169.2  \\ 
709.87842 &  165.7  \\ 
709.88932 &  167.5  \\ 
709.90021 &  175.5  \\ 
710.85013 &  154.4  \\ 
710.86104 &  134.0  \\ 
710.87193 &  145.7  \\ 
710.88283 &  158.2  \\ 
710.89397 &  182.2  \\
768.74817 &   46.4  \\
768.77097 &   55.3  \\
768.79356 &   56.1  \\
\hline
\end{tabular}
\end{table}

\newpage

\setcounter{figure}{0}
    \begin{figure*}
    \centering
    \includegraphics[width=16.2cm]{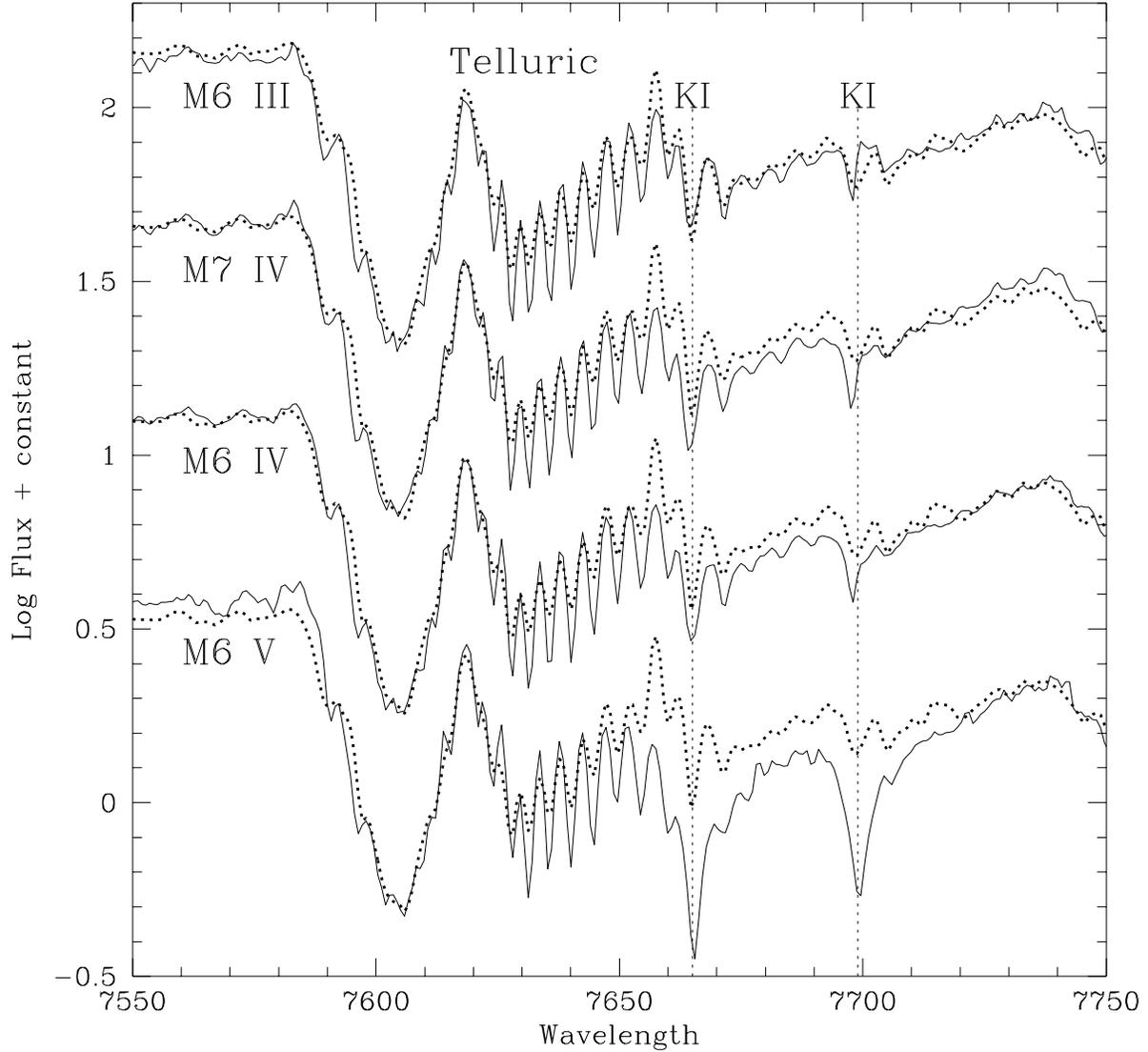}
 \caption{Spectral classification of LS-RCrA~1. Medium resolution
spectra (B\&C instrument,  1200 lines/mm grating) .
The best fit corresponds to luminosity class IV, based on the depth
of the KI 7665\&7699 \AA{ } doublet. 
 The dotted lines corresponds to LS-RCrA~1, whereas the
 solid lines are several 
spectral templates taken with the same setup. 
}
 \end{figure*}

\setcounter{figure}{1}
    \begin{figure*}
    \centering
    \includegraphics[width=15.0cm]{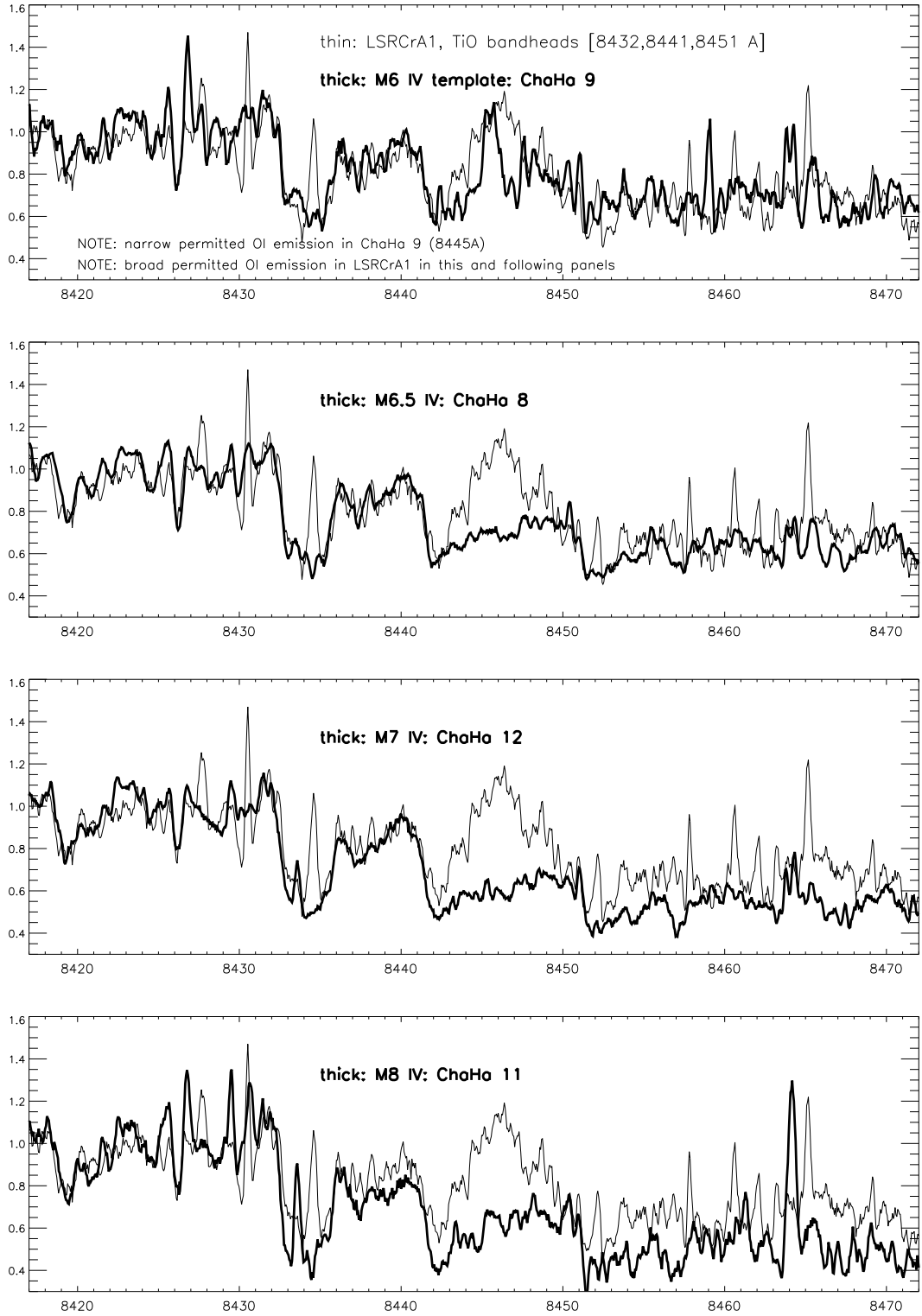}
 \caption{The spectral type estimate of LS-RCrA~1 (thick line)
  based on comparison with PMS  templates from Chameleon (thin lines)
and high-resolution spectra.}
 \end{figure*}

\setcounter{figure}{2}
    \begin{figure*}
    \centering
    \includegraphics[width=16.2cm]{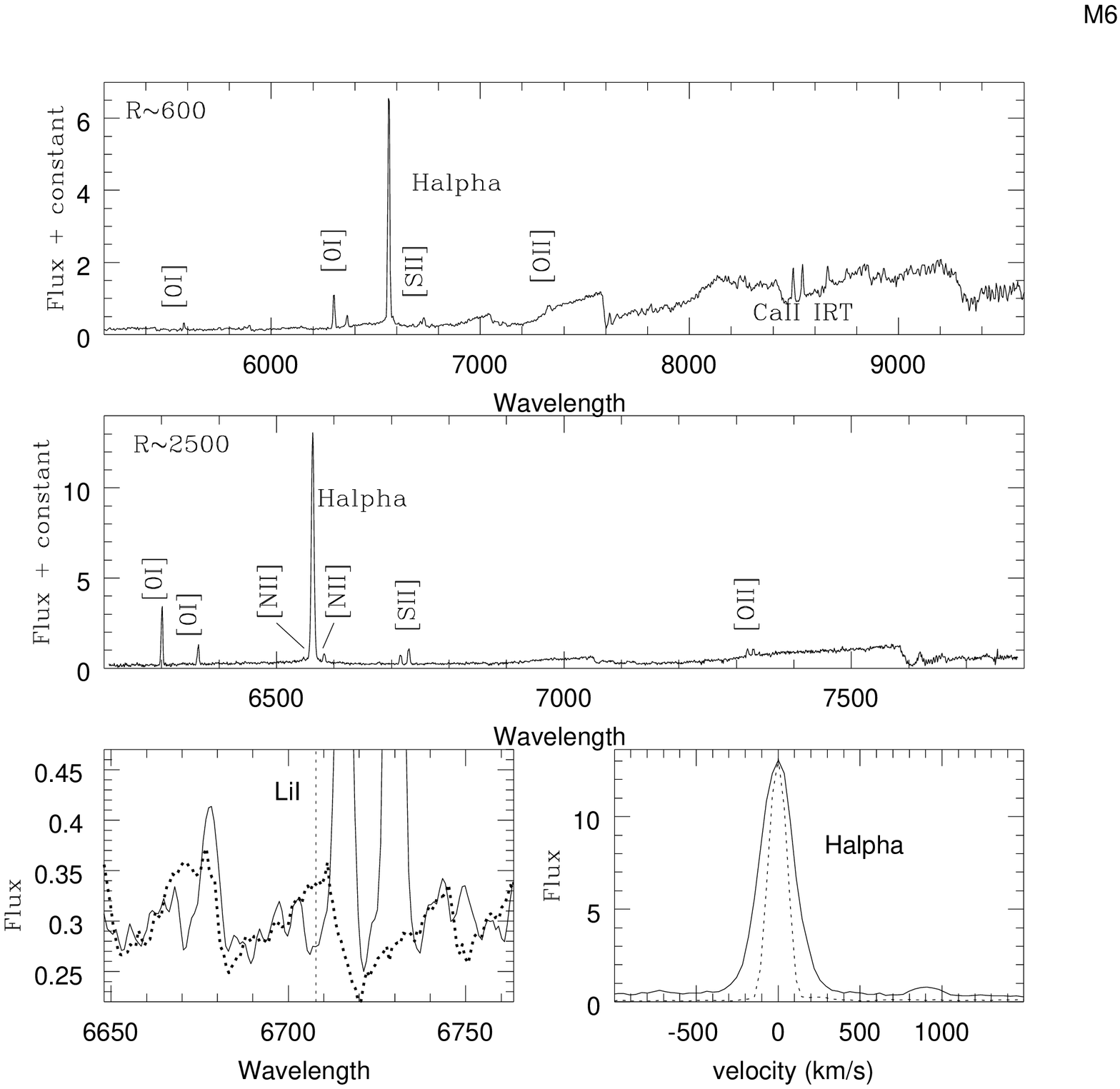}
 \caption{
{\bf a} Upper panel.- Low resolution spectrum. Note the forbidden
 emission lines.  
{\bf b} Middle panel.- Medium resolution spectrum.
{\bf c} Lower panel.- The zoom of the medium resolution spectra
 around H$\alpha$ and [NII]6548\&6581 \AA{} (right);
 and  [SII]6714\&6729 \AA,
 HeI6678 \AA{ } and LiI6708 \AA{} (left).  }
 \end{figure*}

\setcounter{figure}{3}
    \begin{figure*}
    \centering
    \includegraphics[width=15.0cm]{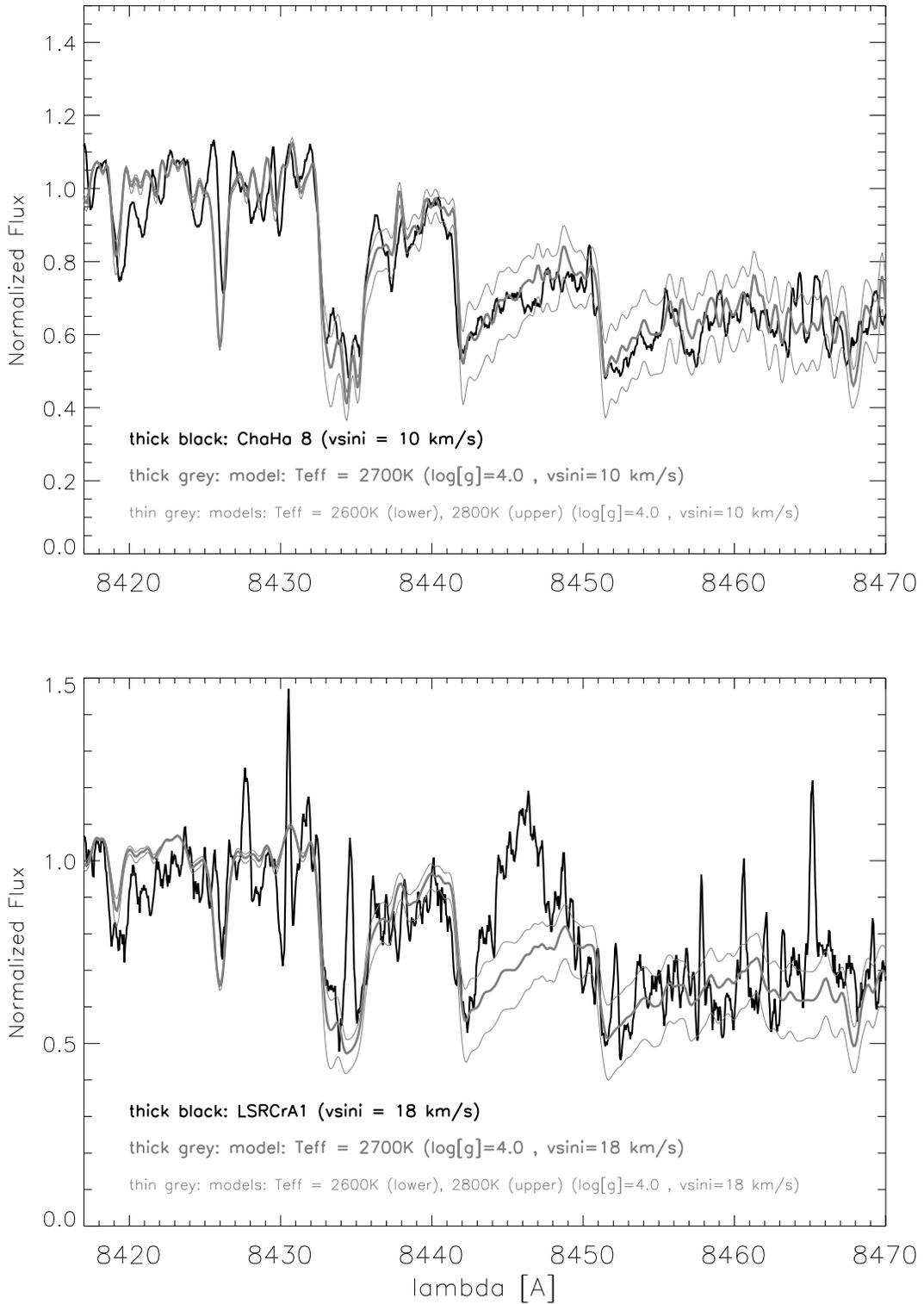}
 \caption{Effective temperature estimate  for ChaHa~8 (top) and
LS-RCrA~1 (bottom) based on high-resolution 
spectra and comparison with synthetic models.  }
 \end{figure*}

\setcounter{figure}{4}
    \begin{figure*}
    \centering
    \includegraphics[width=16.8cm]{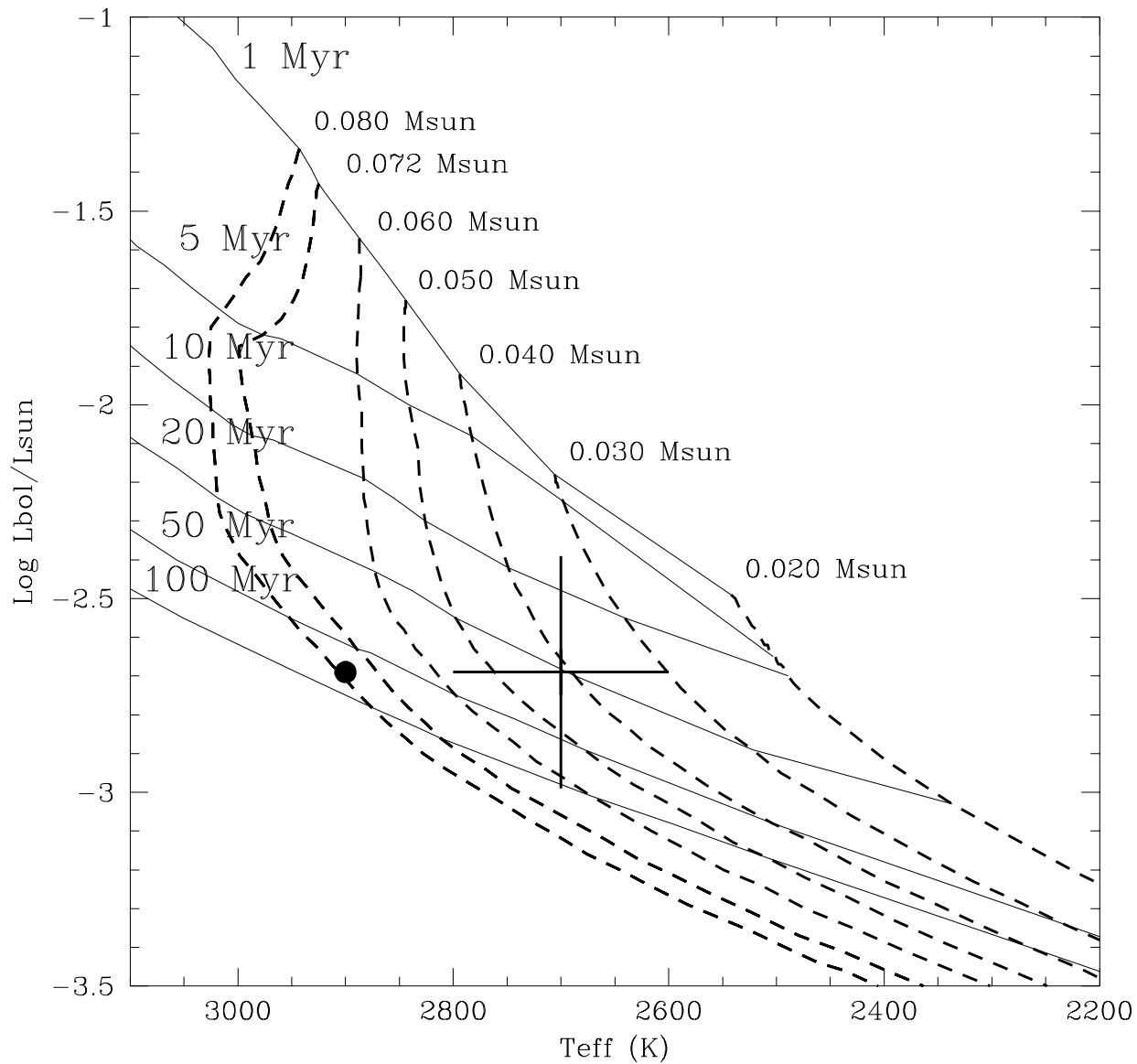}
 \caption{{\bf a} H-R diagram for LS-RCrA~1, located with large 
error-bars (our new estimate) and big solid circle (effective
temperature from Fern\'andez \& Comer\'on 2001).
Isochrones  (1, 5, 10, 20, 50 and 100 Myr) from  Baraffe et al$.$
 (1998) are included in the figure as solid lines.
The thick dotted lines correspond to a evolutionary track for 
0.080, 0.072, 0.060, 0.050, 0.040, 0.030 and 0.020 M$_\odot$.  }
 \end{figure*}

\setcounter{figure}{4}
    \begin{figure*}
    \centering
    \includegraphics[width=16.8cm]{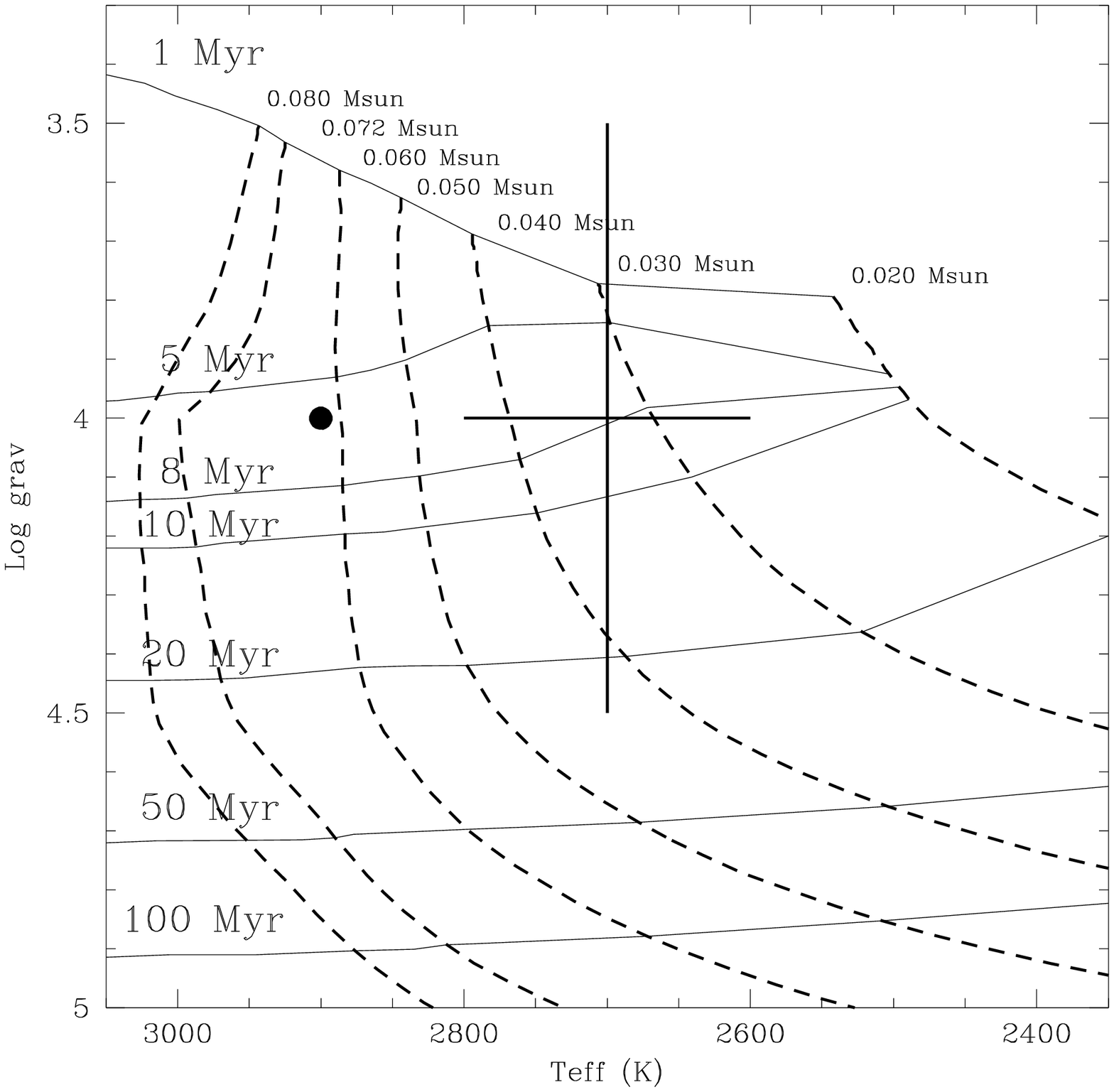}
 \caption{{\bf b} 
Surface  gravity versus de effective temperature. The locations of 
LS-RCrA~1 are indicated with large  error-bars (our new estimate) 
and big solid circle
 (effective temperature from Fern\'andez \& Comer\'on 2001).
Isochrones  (1, 5, 8, 10, 20, 50 and 100 Myr) from  Baraffe et al$.$
 (1998) are included in the figure as solid lines.
The thick dotted lines correspond to a evolutionary track for 
0.080, 0.072, 0.060, 0.050, 0.040, 0.030 and 0.020 M$_\odot$.  }
 \end{figure*}

\setcounter{figure}{5}
    \begin{figure*}
    \centering
    \includegraphics[width=16.2cm]{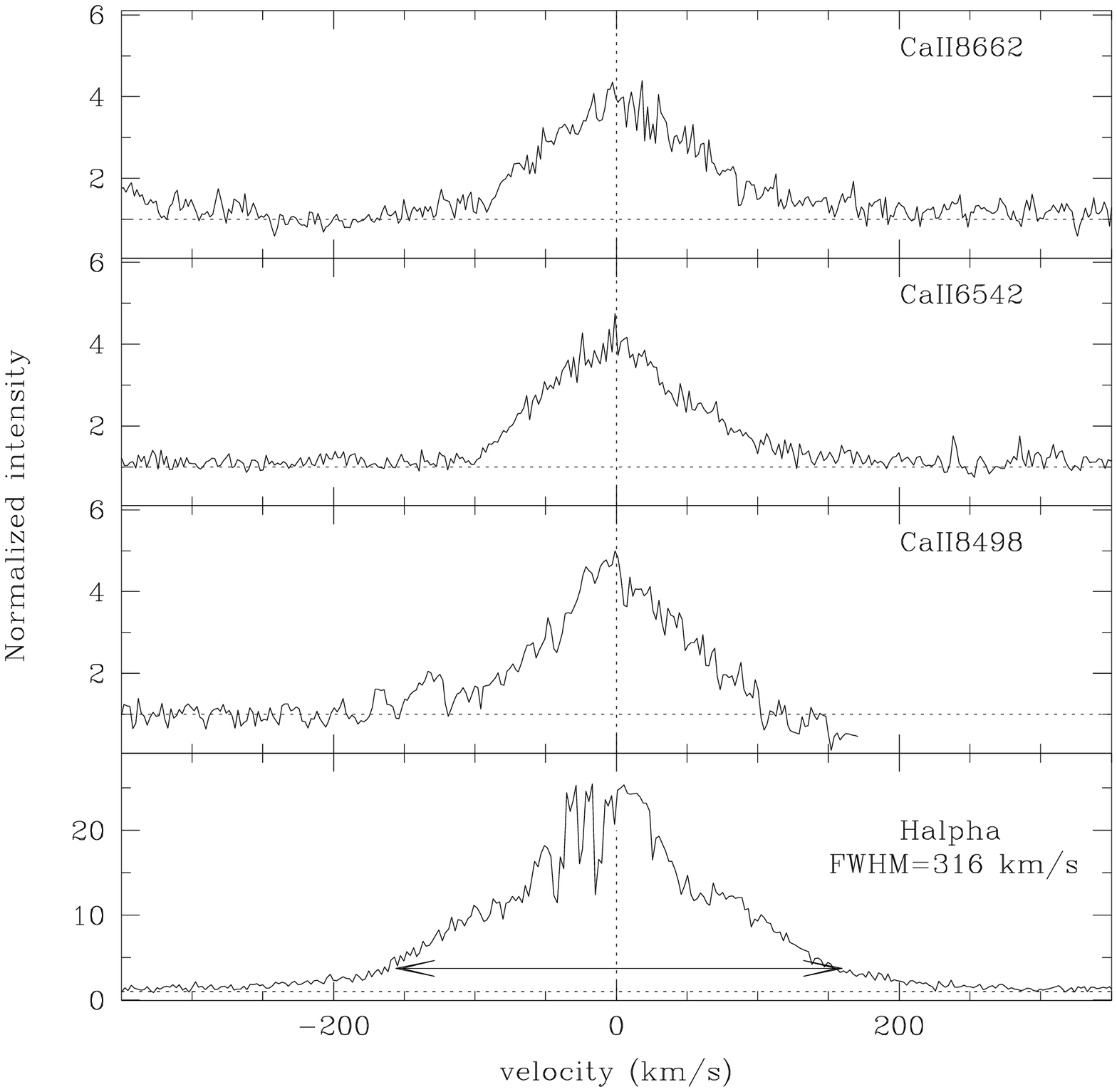}
 \caption{Line profiles for the high resolution data.
{\bf a} Permitted lines.
}
 \end{figure*}

\setcounter{figure}{5}
    \begin{figure*}
    \centering
    \includegraphics[width=16.2cm]{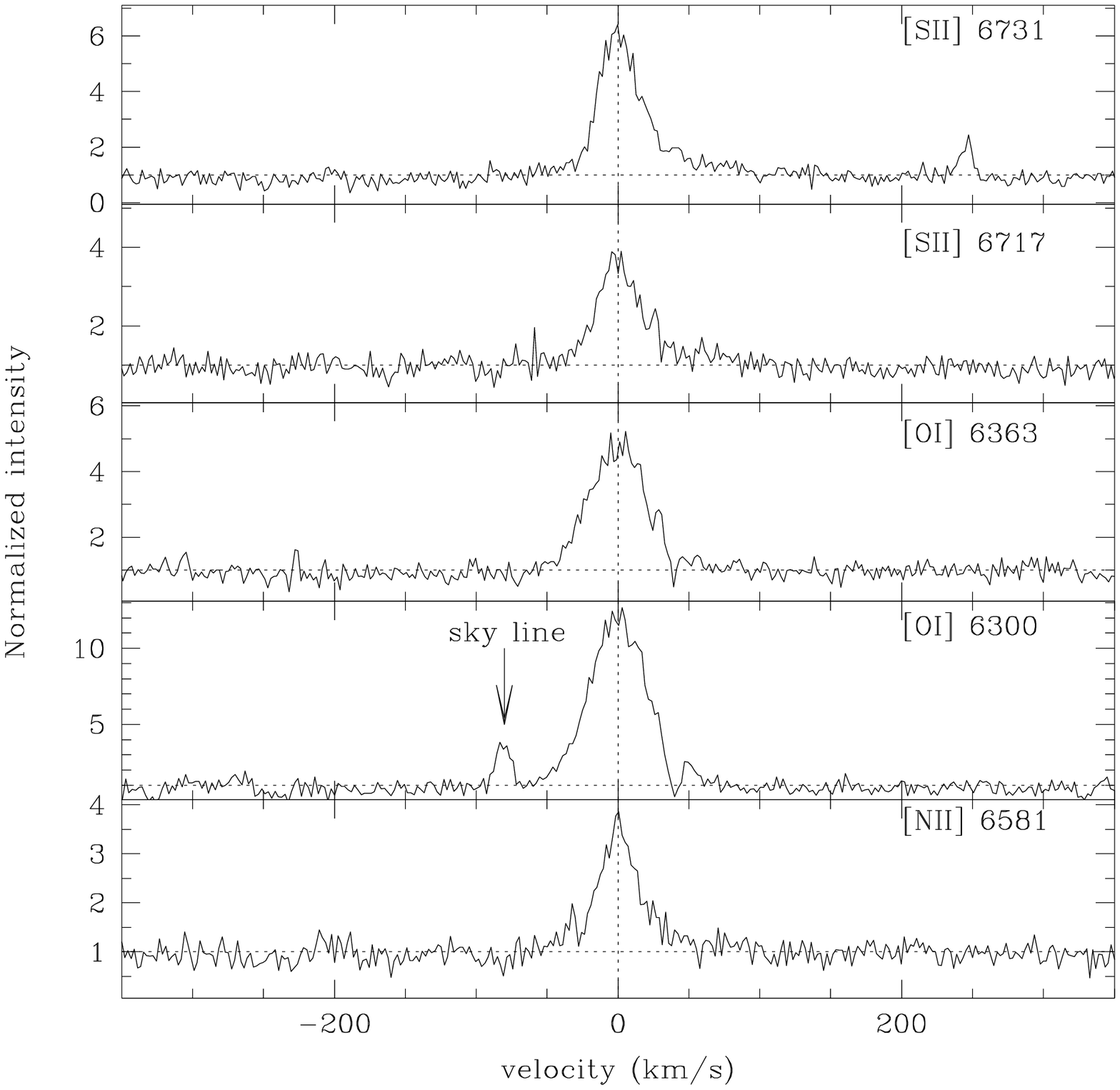}
 \caption{ Line profiles for the high resolution data.
{\bf b} Forbidden lines.
}
 \end{figure*}



\begin{thebibliography}{}

\bibitem[2000]{alencar2000}  
Alencar, S.H.P., \& Basri, G., 
2000, AJ 119, 1881 
 

\bibitem[2001]{allard2001}  
Allard, F., Hauschildt, P.H., Alexander, D.R., Tamanai, A., Sweintzer, A., 
2001, ApJ 556, 357 

 
\bibitem[1998]{baraffe98} 
Baraffe, I., Chabrier, G., Allard, F., 
Hauschildt P. H.,
 1998, A\&A, 337, 403

\bibitem[2002]{baraffe2002} 
Baraffe, I., 
Chabrier, G., Allard, F., Hauschildt, P.H., 
2002, A\&A 382, 563

\bibitem[2001]{barrado2001}  
Barrado y Navascu\'es, D., Bouvier, J., Stauffer, J.R.,  Lodieu, N., McCaughrean, M.J., 
2001, A\&A 396, 813

\bibitem[2002]{barrado2002}  
Barrado y Navascu\'es,  D., Zapatero Osorio,  M.R., Mart\'{\i}n,  E.L., B\'ejar, V.J.S., 
Rebolo,  R., Mundt, R., 
2002, A\&A Letters 393, 85 
 
\bibitem[2003]{barrado2003a}  
Barrado y Navascu\'es, D., B\'ejar, V.J.S., Mundt, R., Mart\'{\i}n, E.L.,  
Rebolo, R., Zapatero Osorio, M.R., Bailer-Jones, C.A.L., 
2003, A\&A, 404, 171 

\bibitem[2003]{barrado2003b}  
Barrado y Navascu\'es, D., Mart\'{\i}n, E.L.,  
 Jayawardhana R., Mohanty S., 
2003, in ``Magnetic fields and star formation'', Kluwer Academic Publishers.
Eds. A.I. G\'omez de Castro, in press.
 
\bibitem[2003]{barrado2003c}  
Barrado y Navascu\'es, D., Mart\'{\i}n, E.L.,  
2003, AJ, accepted.
 
\bibitem[2002]{Bate2002}   
Bate, M.R., Bonnell, I.A., Bromm, V.,  
2002, MNRAS 332, L62 
 
\bibitem[2002]{Bernstein2002} 
 Bernstein, R.A., et al. 2002, Proc. SPIE 4841, 1694

\bibitem[1997]{burrow1997} 
Burrow A.,  Marley  M., Hubbard  W.B.,
 Lunine  J.I., Guillot T., Saumon  D.,
 Freedman  R., Sudarsky D., Sharp C.,
1997, ApJ 491, 856

\bibitem[2001]{Carpenter2001} 
Carpenter, J.M.,
2001, AJ 121, 2851

\bibitem[1998]{Casey1998} 
Casey, B.W:, Mathieu, R.D:, Vaz, L.P.R., Anderson, J., Suntzeff, N.B., 
1998, AJ 115, 1617.

\bibitem[Comeron et al. (2003)]{comeron2003} 
Comer\'on, F., Fern\'ndez, M., Baraffe, I.,
 Neuh\"user, R., Kaas, A.A.,
2003, A\&A 406, 1001

\bibitem[Cutri et al. (2003)]{cutri2003} 
Cutri, R.M., et al$.$
2003, ``2MASS All-Sky Catalog of Point Sources'',
University of Massachusetts and Infrared Processing 
and Analysis Center, (IPAC/California Institute of Technology).

\bibitem[1997]{DAntona1994}
D'Antona, F., \& Mazzitelli, I. 
1994 ApJ Suppl. 90, 467

\bibitem[fernandez2001]{fernandez2001} 
Fern\'andez, M., Comer\'on, F.,
2001, A\&A 380, 264 [F\&C01]

\bibitem[1996]{Green1996} 
Greene, T.P., Lada, C.J.,
1996, AJ 112, 2184

\bibitem[2003]{hanuschik2003} 
Hanuschik, R.W., 
2003, A\&A, 407, 1157

\bibitem[1997]{hartmann1997} 
Hartmann, L., Cassen, P., Kenyon, S.J.,
1997, ApJ 475, 770


\bibitem[2002]{jayawardhana2002a}  
Jayawardhana, R., Mohanty, S., Basri, G., 
2002, ApJ Letters 578, 141  

\bibitem[2002]{jayawardhana2002b}
Jayawardhana, R., Luhman, K.L., D'Alessio, P., \& Stauffer, J. 2002, 
ApJ, 571, L51
 
\bibitem[2003]{jayawardhana2003a}  
Jayawardhana, R., Mohanty, S., Basri, G., 
2003, ApJ 592, 282
 
\bibitem[2003]{jayawardhana2003b} 
Jayawardhana, R., Ardila, D.R., Stelzer, B., Haisch, K.E.,  
2003, AJ 126, 1515

\bibitem[1992]{Leggett1992}   
Leggett, S., 1992 ApJ SS 82, 351


\bibitem[2000]{leggett2000} 
Leggett, S.K.,  Allard, F., Dahn, C.,  
Hauschildt, P. H., Kerr, T. H., Rayner, J. 
2000, ApJ 535, 965 
 
\bibitem[2001]{leggett2001} 
Leggett, S.K., Allard  F., Geballe  T.R.,
 Hauschildt  P.H., Schweitzer  A., 
2001, ApJ 548, 908.

\bibitem[2002]{leggett2002} 
Leggett, S.K., Golimowski, D.A., Fan, X., et al$.$ 
2002, ApJ 564, 452 

\bibitem[2003]{liu2003} 
Liu, M.C., Najita, J., Tokunaga, A.T., 
2003, ApJ 585, 372 
 
\bibitem[1999]{luhman1999} 
  Luhman,  K.L. 
 1999, ApJ, 525, 466 

\bibitem[2003]{luhman2003b}  
Luhman,  K.L.,  Stauffer, J.R., Muench, A.A., Rieke, G.H., Lada, E.A.,  
Bouvier, J., Lada, C.J., 
2003, ApJ 593, 1093

\bibitem[2003]{luhman2003a} 
Luhman,  K.L., Brice\~no, C., Stauffer, J.R., Hartmann, L.,  
Barrado y Navascu\'es, D., Caldwell, N.,  
2003a, ApJ, 590, 348 
 
\bibitem[1981]{Marraco1981} 
Marraco,  H.G., Rydgren, A.E., 
1981, AJ 86, 62


\bibitem[2003]{mohanty2003a} 
Mohanty, S., Basri, G.,  
2003, ApJ 583, 451 

\bibitem[2003]{Mohanty2003b} 
Mohanty, S., Jayawardhana, R., Barrado y Navascu\'es, D.,
2003, ApJ Letters 593, 109 
 
\bibitem[2001]{muench2001} 
Muench, A.A., Alves, J., Lada, C.J., Lada, E.A.,  
2001 ApJ Letters 558, 51 

\bibitem[1998]{muzerolle1998a} 
  Muzerolle, J.,  Calvet, N., Hartmann, L., 
1998a, ApJ 492,  743

\bibitem[1998]{muzerolle1998b} 
  Muzerolle, J.,  Hartmann, L., Calvet,N. 
1998b, AJ, 116, 455


\bibitem[2001]{muzerolle2001} 
  Muzerolle, J.,  Calvet, N., Hartmann, L., 
2001, ApJ 550, 944

\bibitem[2003]{muzerolle2003} 
Muzerolle, J., Hillenbrand, L., Calvet, N., Brice\~no, C., Hartmann, L. 
2003, ApJ, 592, 266

\bibitem[2002]{Natta2002} 
Natta, A., Testi, L., Comer\'on, F., 
 Oliva, E., D'Antona, F., Baffa, C., 
 Comoretto, G., Gennari, S., 
2002, A\&A 393, 597 
 
\bibitem[2000]{Neuhauser2000} 
Neuha\"user, R., Walter, F.M., Covino, E., Alcal\'a, J.M., 
Wolk, S.J., Frink, S., 
Guillout, M.F:, Sterzik, M.F., Comer\'on, F. 
2000, A\&AS 146, 323 [N00]

\bibitem[2002]{padoan2002} 
Padoan, P., \& Nordlund, \AA., 
2002, ApJ 576, 870 
 
\bibitem[2001]{reipurth2001} 
Reipurth, B., \& Clarke, C., 
2001, AJ 122, 432 

\bibitem[1985]{rieke1985}  
Rieke G.H.  \& Lebofsky M.J.,
1985, ApJ 288, 618

\bibitem[2003]{segransan}
S\'egransan  D., Kervella  P., Forveille T.,  Queloz D.,
2003, A\&A Letters 397, 5

\bibitem[2000]{Siess2000}
Siess L., Dufour E., Forestini M., 2000, A\&A 358, 593

\bibitem[1999]{stauffer1999} 
Stauffer, J.R., Barrado y Navascu\'es, D., Bouvier, J., et al.
1999, ApJ 527, 219 

\bibitem[2003]{White2003} 
White, R.J.,  Basri, G.,
2003, ApJ 582, 1109

\bibitem[1997]{Wiking1997} 
Wilking, B.A.,  McCaughrean, M.J., Burton, M.G.,
 Giblin, T., Rayner, J.T., Zinnecker, H.,  
1997, AJ 114, 2029.

\bibitem[1997]{zapatero1997} 
Zapatero Osorio, M.R., Mart\'{\i}n, E.L., Rebolo, R.,   
1997, A\&A 323, 105


\end{thebibliography}
\end{document}